\documentclass[twoside,11pt]{article}  
\usepackage{CJK} 
\usepackage{indentfirst} 
\usepackage{bm} 
\usepackage{graphicx} 
\usepackage{epsfig} 
\usepackage{amsmath} 

\begin{document} 

\begin{CJK*}{GBK}{song}  



\begin{center}
\LARGE\bf Thermal entanglement in the mixed
three spin XXZ Heisenberg Model on a triangular cell   
\end{center}

\begin{center}  
S. Deniz Han and Ekrem Aydiner
\end{center}

\begin{center}  
\begin{small} \sl
Department of Physics, \.{I}stanbul University, Tr-34134,
\.{I}stanbul, Turkey
\end{small}
\end{center}

\begin{center}  
\small (Received X XX XXXX; revised manuscript received X XX XXXX)
\end{center}

\vspace*{2mm}

\begin{center}  
\begin{minipage}{15.5cm}
\parindent 20pt\small
We numerically investigate thermal entanglement of the spins (1/2,1)
and (1/2,1/2) in the three-mixed (1/2,1,1/2) anisotropic Heisenberg
XXZ spin system on a simple triangular cell under an inhomogeneous
magnetic field. We show that the external magnetic field induces
strong plateau formation in pairwise thermal entanglement for fixed
parameters of Hamiltonian in the case of the ferromagnetic and
anti-ferromagnetic interactions. We also observe an unexpected
critical point at finite temperature in the thermal entanglement of
the spins (1/2,1) for antiferromagnetic case while the entanglement
of the spins (1/2,1) for ferromagnetic case and the entanglement of
the spins (1/2,1/2) for both ferromagnetic and antiferromagnetic
cases almost decays exponentially to zero with increasing
temperature. The critical point in entanglement of the spins (1/2,1)
for antiferromagnetic case may be signature of the quantum phase
transition at finite temperature.

\end{minipage}
\end{center}

\begin{center}  
\begin{minipage}{15.5cm}
\begin{minipage}[t]{2.3cm}{\bf Keywords:} \end{minipage}
\begin{minipage}[t]{13.1cm}
Quantum entanglement, quantum phase transition,negativity

\end{minipage}\par\vglue8pt
{\bf PACC: } 03.65.Ud, 64.70.Tg

\end{minipage}
\end{center}

\footnotetext[1]{Project supported by Istanbul University (under
grant numbers: 19240 and 28432).} \footnotetext[2]{Corresponding
author. E-mail: ekrem.aydiner@istanbul.edu.tr}

\section{Introduction}

It has been known that the quantum entanglement in spin systems
plays crucial role in physics such as in quantum information
\cite{Nielsen}, quantum computation
\cite{Schumacher,Loss,Burkard,Imamoglu,Platzman,Raussendorf},
quantum teleportation \cite{Bennett1}, superdense coding
\cite{Bennett2}, quantum communication \cite{Divincenzo,Bose,Kay},
quantum perfect state transfer \cite{Christandl}, quantum cryptology
\cite{Ekert,Deutsch} and quantum computational speed-ups
\cite{Shor,Grover}. Potential applications of spin-spin entanglement
in these fields have stimulated researches on methods to quantify
and control it. Therefore, many detailed theoretical and
experimental studies have been performed to understand the quantum
entanglement behavior in two and more qubits which consist of
spin-half, mixed-spin or higher Heisenberg (XX, XY, XXZ and XYZ) and
Ising spin systems.

Recently it has been shown that the molecular spin ring, triangular
spin cell or different spin lattice configuration have an important
potential in computation and information sciences as a entanglement
source. On the other hand it is known that the entanglement in a
mixed spin systems with a certain geometry may produce rich behavior
more than simple i.e., two or more identical spin systems. For
instance, the three-qubit entangled states have been shown to
possess advantages over the two-qubit states in quantum
teleportation \cite{Karlsson}, dense coding \cite{Hao} and quantum
coloning \cite{Brub}. However, there are limited works on the
mixed-spin systems with different lattice (triangular or square)
configuration in the literature \cite{Sun,Shun}. It still has
importance to clarify of entanglement behavior of different spin
systems which consist of mixed spins with different lattice
configuration. Therefore, in this paper, we focus on entanglement of
three-spin (1/2,1,1/2) anisotropic Heisenberg XXZ system on a simple
triangular cell under an inhomogeneous magnetic field at
equilibrium. We will show that the external magnetic field induces
strong plateau formation in thermal entanglement in the
ferromagnetic (F) and antiferromagnetic (AF) XXZ model and
unexpected critical point in entanglement occurs at finite
temperature for AF interaction.

The paper is organized as follows. In Section 2, the model and
method are introduced. In Section 3, we numerically obtain the
pairwise thermal entanglement between qubits in model and we discuss
quantum phase transition. Conclusion is given in Section 4.

\section{Model and Method}

The three-spin (1/2,1,1/2) anisotropic Heisenberg XXZ system on a
simple triangular cell schematically illustrated in Fig.\,(1) with
interacting coupling. Spin-1/2 and spin-1 are represented by black
circles and white circle, respectively. As it can be seen from
Fig.\,(1) that triangular cell with spins can be mapped to one
dimensional system. Now, one-dimensional spin system can be
considered as an qutrit-qubits system. The Hamiltonian of the
three-spin (1/2,1,1/2) anisotropic Heisenberg XXZ system on a
one-dimensional lattice under an inhomogeneous magnetic field is
given as
\begin{eqnarray}
H=\sum_{i=1}^{3}[J(s_{i}^{x}s_{i+1}^{x}+s_{i}^{y}s_{i+1}^{y})+\gamma
(s_{i}^{z}s_{i+1}^{z})] +(B+b)s_{1}^{z}+Bs_{2}^{z}+(B-b)s_{3}^{z} \
.
\end{eqnarray}
In this Hamiltonian, $s_{i}^{x,y,z}$ ($i=1,2,3$) represents spin
operators of the spin-1/2 and spin-1 components and the exchange
interaction parameters are chosen as $J=J_{1}=J_{2}\neq J_{3}$
($J_{3}=\gamma $). Here $J$ denotes the exchange interaction between
the nearest-neighboring spin pairs, $\gamma$ is the anisotropy
parameter ($-1\leq \gamma \leq 1$); $B$ and $b$ are strengths of
homogenous and stagger magnetic field, respectively. The chain has F
interaction for $J<0$ and AF interaction for $J>0$. The periodic
boundary conditions satisfy $s_{3}^{x,y}=s_{1}^{x,y}$.
\begin{figure} 
\centering{{\includegraphics[width=4.5cm, height=8cm]{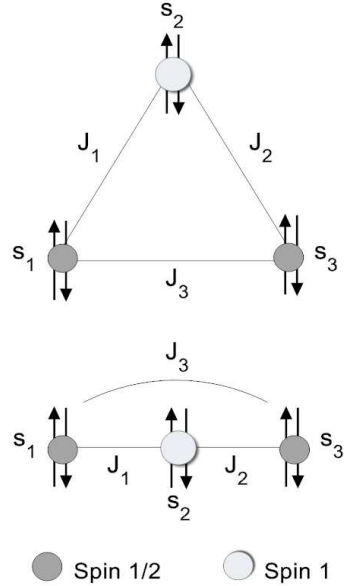}}
\caption{Triangle lattice representation of the mixed spin system
for the three spins with coupling constants.}}
\end{figure}

In order to obtain thermal entanglement for Hamiltonian (1) we use
the concept of the negativity. It is known that entanglement in
spin-half systems can be measured using by concurrence which is
applicable to an arbitrary state of two spin halves \cite{Wootters}.
However, for systems which consist of the higher spins the thermal
entanglement of the spin system can be analyzed using by negativity.
The negativity which is quantitative version of the Peres-Horodecki
criterion \cite{Peres,Horodecki} has been proposed by Vidal and
Werner \cite{Vidal}. They presented that entanglement for higher
spins can be computed efficiently using by negativity, and the
negativity does not increase under local manipulations of the
system. The negativity is defined, depending on density operator
$\rho \left( T\right)$, by
\begin{equation}
N\left( \rho\right)=\sum_{i}\left\vert \mu _{i}\right\vert ,
\end{equation}
where $\mu _{i}$ is the negative eigenvalue of partial transpose
$\rho \left( T\right)^{t_{1}}$. Here $t_{1}$ and $\rho \left(
T\right) $ are the partial transpose with respect to the first
system and density operator, respectively. The density operator
$\rho \left( T\right) =\exp \left( -H/kT\right) /Z$ represents the
state of the system at thermal equilibrium where $Z=Tr\left( \exp
\left( -H/kT\right) \right) $ is the partition function, $H$ is the
Hamiltonian, $T$ is the temperature and $k$ is Boltzmann's constant
which we take equal to $1$ for the sake of simplicity. Also Eq.\,(1)
can be written related to the trace norm of $\rho ^{t_{1}}$ as
\begin{equation}
N\left( \rho\right)=\frac{\left\Vert \rho \left( t\right)
^{t_{1}}\right\Vert -1}{2} \ ,
\end{equation}
where the trace norm of $\rho ^{t_{1}}$ is equal to the sum of the
absolute values of the eigenvalues of $\rho ^{t_{1}}$.

The entanglement between each spin-pairs in this model can be
obtained separately. Therefore the negativity $N_{12}$ and $N_{13}$
correspond to entanglement between spins (1/2,1) and (1/2,1/2),
respectively. The computation of the negativities $N_{12}$ and
$N_{13}$ require to known the partial transpose of reduced density
matrices which is special form of the density matrix $\rho$ of
Hamiltonian (1). Therefore, to investigate the thermal entanglement
of this system, the simple procedure is as follows. In the first
step, the Hamiltonian is written in the Hilbert space as a
block-diagonal form using above basis. Thus density matrix $\rho$ of
Hamiltonian (1) can be obtained in terms of eigenvalues of the a
block-diagonal form of the Hamiltonian. In the second step, reduced
density matrices of $\rho$ can be obtained after tracing out with
respect to related any spin, then the pairwise negativities
($N_{12}$ and $N_{13}$) are computed according Eq.\,(2) by using the
basis vectors of the three-spin system defined in Eq.\,(1), which
are $\{$$\left\vert -1/2,-1,-1/2\right\rangle$, $\left\vert
1/2,-1,-1/2\right\rangle$, $\left\vert -1/2,-1,1/2\right\rangle $,
$\left\vert 1/2,-1,1/2\right\rangle$, $\left\vert
-1/2,0,-1/2\right\rangle$, $\left\vert 1/2,0,-1/2\right\rangle$,
$\left\vert -1/2,0,1/2\right\rangle$, $\left\vert
1/2,0,1/2\right\rangle$, $\left\vert -1/2,1,-1/2\right\rangle$,
$\left\vert 1/2,1,-1/2\right\rangle$, $\left\vert
-1/2,1,1/2\right\rangle$, $\left\vert 1/2,1,1/2\right\rangle$$\}$,
where $\left\vert s_{1},s_{2},s_{3}\right\rangle $ is the eigenstate
of $s_{1}^{z},s_{2}^{z},s_{3}^{z}$ with corresponding eigenvalues
given by $s_{1},s_{2},s_{3}$, respectively.

In the light of the above mentioned explanations we give partial
transpose $\rho _{12}^{t_{1}}$ and $\rho _{13}^{t_{1}}$ of reduced
density matrices with elements below. For example; after tracing out
the third spin-half system, the partial transpose $\rho
_{12}^{t_{1}}$ of the reduced density matrix $\rho _{12}$ is written
in the form
\begin{equation}
\rho _{12}^{t_{1}}=\frac{1}{Z}\left(
\begin{array}{cccccc}
a_{11} & a_{12} & 0 & 0 & 0 & 0 \\
a_{21} & a_{22} & 0 & 0 & 0 & 0 \\
0 & 0 & a_{33} & 0 & 0 & 0 \\
0 & 0 & 0 & a_{44} & 0 & 0 \\
0 & 0 & 0 & 0 & a_{55} & a_{56} \\
0 & 0 & 0 & 0 & a_{65} & a_{66}%
\end{array}%
\right)
\end{equation}
where $a_{12}=a_{21}$ and $a_{56}=a_{65}$. Hence the negativity
$N_{12}$ for the (1/2,1) system is given by
\begin{eqnarray}
N(\rho _{12})=\frac{1}{2}\max \left[ 0,\sqrt{(a_{11}-a_{22})^{2}+4a_{21}^{2}%
}-a_{11}-a_{22}\right]\nonumber\\
+\frac{1}{2}\max \left[ 0,\sqrt{%
(a_{55}-a_{66})^{2}+4a_{65}^{2}}-a_{55}-a_{66}\right]
\end{eqnarray}
On the other hand, after tracing out the second spin-1, the partial
transpose $\rho _{13}^{t_{1}}$ of the reduced density matrix $\rho
_{13}$ is written in the form
\begin{equation}
\rho _{13}^{t_{1}}=\frac{1}{Z}\left(
\begin{array}{cccc}
a_{11}^{\prime} & 0 & 0 & a_{41}^{\prime} \\
0 & a_{22}^{\prime} & 0 & 0 \\
0 & 0 & a_{33}^{\prime} & 0 \\
a_{14}^{\prime} & 0 & 0 & a_{44}^{\prime}%
\end{array}%
\right)
\end{equation}
where $a_{11}^{\prime}=a_{44}^{\prime}$,
$a_{14}^{\prime}=a_{41}^{\prime}$ and
$a_{22}^{\prime}=a_{33}^{\prime}$. Finally the negativity $N_{13}$
for the (1/2,1/2) system is given depend on matrix elements of
Eq.\,(6) by
\begin{equation}
N(\rho _{13})=\max \left[ 0,\left\vert a_{41}^{\prime}\right\vert
-a_{11}^{\prime}\right] \ .
\end{equation}
However it is very hard to write the elements of the matrices $\rho
_{12}^{t_{1}}$ and $\rho _{13}^{t_{1}}$ in explicit form. Therefore
computing numerically the elements of the matrices in Eqs.(4) and
(6), we obtain the negativity $N(\rho_{12})$ and $N(\rho_{13})$ in
Eqs.(5) and (7), respectively.

\section{Numerical Results}

In this section we present our numerical results of thermal
entanglement between spins in three-spin (1/2,1,1/2) anisotropic
Heisenberg XXZ system under an inhomogeneous magnetic field. The
negativities  $N_{12}$ and $N_{13}$ for F and AF interactions are
present in subsections, respectively.

\subsection{Thermal Entanglement between spins-(1/2,1) in spin-(1/2,1,1/2) system}

To witness entanglement between spins (1/2,1) in three-spin
(1/2,1,1/2) anisotropic Heisenberg XXZ system under an inhomogeneous
magnetic field for F and AF interactions, the negativity $N_{12}$ in
Eq.\,(5) is obtained after computing numerically elements of the
matrix $\rho _{12}^{t_{1}}$. Obtained results are given below.

The magnetic field dependence of the negativity $N_{12}$ for stagger
magnetic fields $b=0$ and $b=0.5$ at fixed anisotropy parameter
$\gamma=0.5$ and temperature $T=0.01$ are shown in Fig.\,2 (a) for F
case  $J=-1$,  and in Fig.\,2 (b) for AF case $J=1$, respectively.
As it can be seen from Fig.\,2 that typically two plateaus appear in
negativity curve depend on external magnetic field $B$ when other
parameters of Hamiltonian are fixed in the case of F and AF
interactions.
\begin{figure}  
\centering{{\includegraphics[width=9cm, height=6cm]{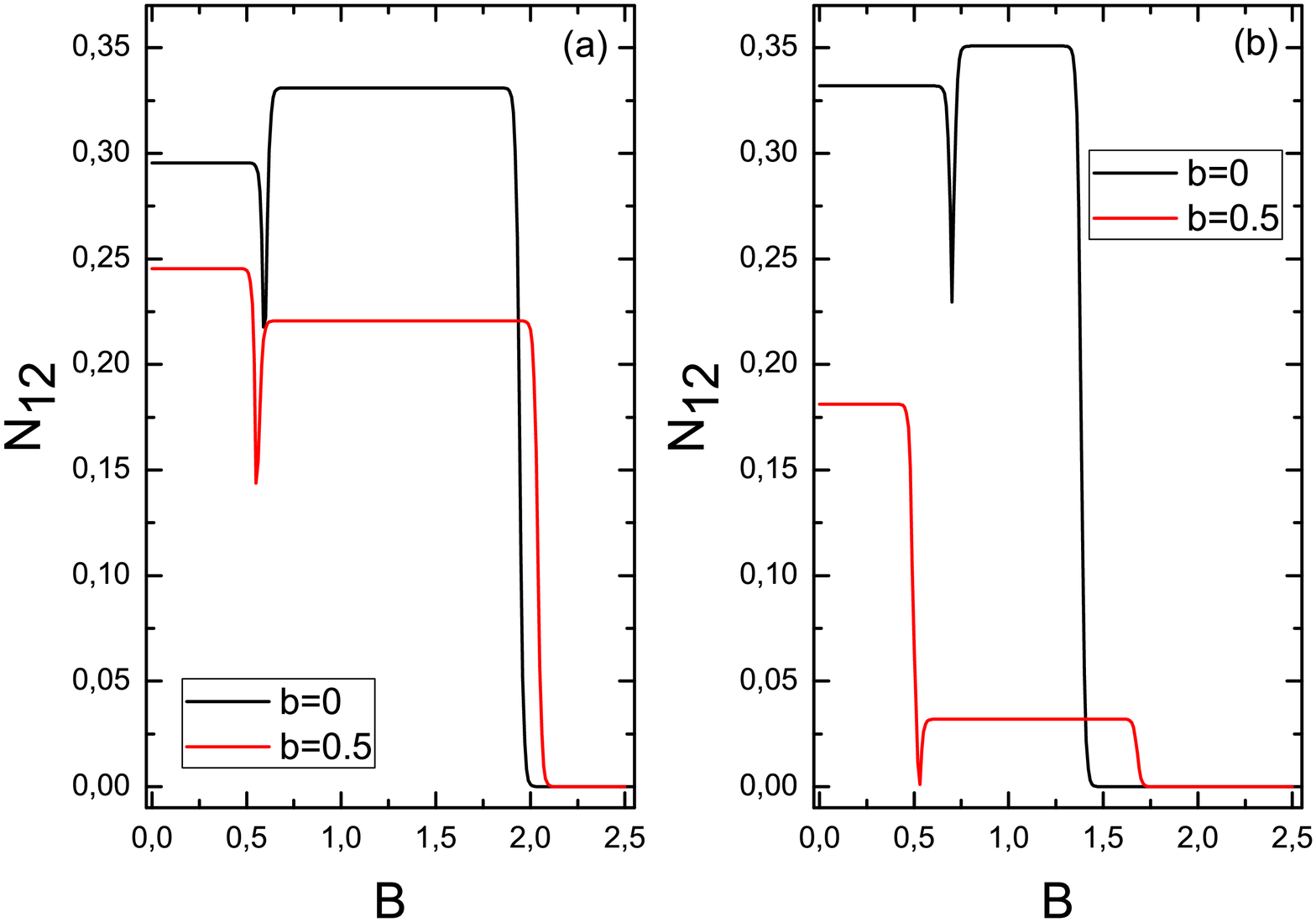}}
\caption{Magnetic field dependence of the $N_{12}$ for arbitrary
anisotropy and staggered magnetic field values at very low
temperature (a) F case $J=-1$ (b) AF case $J=1$. In (a) and (b) the
parameters are set as $\gamma=0.5$, $T=0.01$, $b=0$ and $b=0.5$.}}
\end{figure}

Transition from the first plateau to the second plateau occur at a
critical magnetic field $B_{c1}$ where the negativity $N_{12}$ has
an singularity and, on the other hand, second entanglement plateau
suddenly drop to zero at the second critical magnetic field
$B_{c2}$. The stagger magnetic field does not play role as a plateau
mechanism however it changes the width and height of the
entanglement plateaus. To explain entanglement plateau in the
negativity curve we can apply an analogy between entanglement and
magnetization since entanglement is also a qualitative measure of
the correlations between spins as well magnetization. Haldane
predicted \cite{Haldane} that the energy gap between singlet ground
states and the first excited triplet states in spin systems can
originate from frustration, dimerization, single-ion anisotropy or
periodic fields and leads to plateau behavior in magnetization under
external magnetic field. Hence we can conclude that the similar
energy gap between singlet ground states and the first excited
triplet states in three-spin (1/2,1,1/2) anisotropic Heisenberg XXZ
system may leads to plateau behavior in entanglement.
\begin{figure}  
\centering{{\includegraphics[width=9cm, height=6cm]{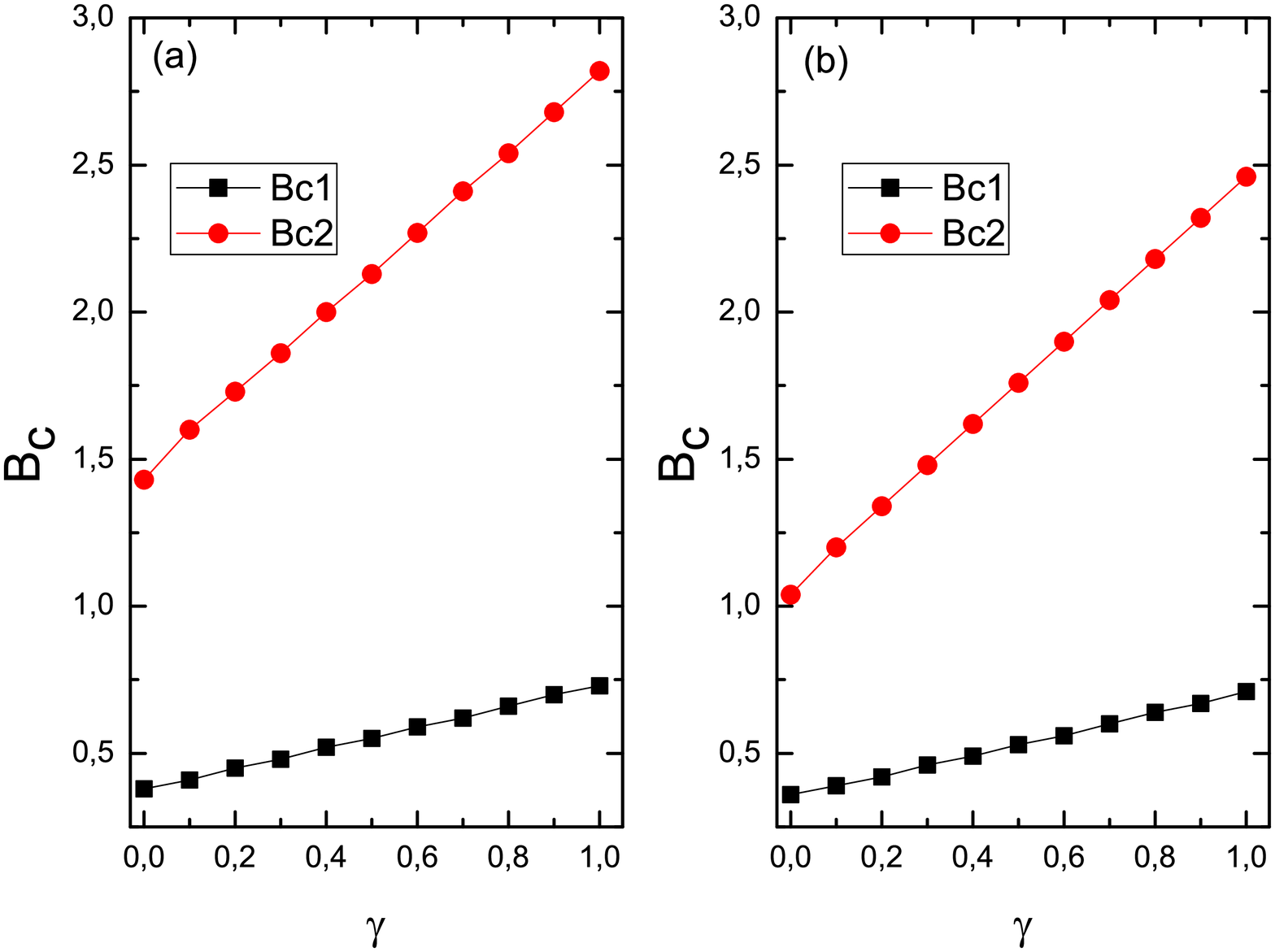}}
\caption{Anisotropy dependence of the $N_{12}$ in the F and AF XXZ
systems for $B=1$, $b=0.5$ and $T=0.01$ in (a) F case ie., $J=-1$,
(b) AF case i.e., $J=1$.}}
\end{figure}

Critical behavior in negativity $N_{12}$ clearly depends on all
parameters of Hamiltonian (3). However anisotropy and temperature
play significant role like external magnetic field on the
entanglement the behavior in the spin systems. Therefore, to
investigate the anisotropy dependence of the critical behavior of
negativity $N_{12}$, critical $B_{c}$ is plotted versus anisotropy
values in Fig.\,3 for F and AF case at fixed temperature and
staggered field values. As it can be seen from Fig.\,3 (a) and (b)
that critical $B_{c1}$ and $B_{c2}$ linearly increase with
increasing anisotropy $\gamma$ value for fixed temperature and
staggered field values. The width of the plateaus also linearly
decreases when anisotropy decreases from one to zero. For
$\gamma>0$, three-spin (1/2,1,1/2) F and AF triangle behaves as XXZ
system and when $\gamma=0$ the system reduces to XX system.
Appearance of plateau at $\gamma=0$ in Fig.\,3 is type that plateau
behavior in entanglement occurs in three-spin (1/2,1,1/2) XX
Heisenberg spin system.

\begin{figure} 
\centering{{\includegraphics[width=8cm, height=6cm]{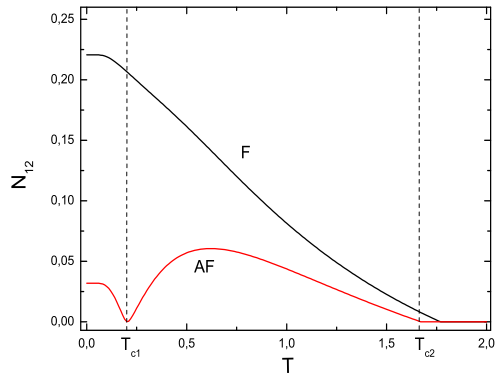}}
\caption{The $N_{12}$ as a function of temperature $T$ at $B=1$,
$b=0.5$ and $\gamma=0.5$ for F case ie., $J=-1$ and AF case i.e.,
$J=1$.}}
\end{figure}
To investigate temperature dependence of the entanglement between
spins, the negativities $N_{12}$ for fixed $J=\pm1$, $\gamma=0.5$,
$B=1$ and $b=0.5$ values are plotted as a function of temperature in
Fig.\,4. It can be seen from figure that, in case of F ($J=-1$), the
negativity $N_{12}$ decreases monotonically with temperature $T$ and
finally reaches to zero at a critical temperature $T_{c}$. Thermal
entanglement in F case is more sensitive to the temperature than
that in AF, however, the negativity $N_{12}$ in the case of AF
($J=1$) exhibits very interesting behavior. Indeed, the negativity
for AF firstly monotonically decrease with temperature $T$ and it
reaches to zero at a critical $T_{c1}$. However the negativity again
shows a smooth revival until a maximum value, due to the optimal of
all eigenstate in the system, then collapses gradually zero at a
second critical $T_{c2}$ when temperature is increased. This
critical behavior at $T_{c1}$ in the negativity $N_{12}$ for AF case
may be signature of the quantum phase transition at finite
temperature. We notice that quantum phase transition is purely
driven by quantum fluctuations where the de Broglie wavelength is
greater than the correlation length of thermal fluctuations
\cite{Werlang2,Wu} and is generally expected at absolute zero
temperature. In a recent studies quantum phase transition in
entanglement of two or more qubits have been observed at absolute
zero temperature, for example, in
Refs.\,\cite{Hu,Wang,Abliz,Zhang,ChaoLi}. However quantum phase
transition can also occur near absolute temperature where thermal
fluctuations are negligible \cite{Werlang2,Werlang1,Werlang3}. At
finite temperature if the system resist to thermal fluctuation due
to entanglement between particles, quantum fluctuations may survive
in the system and it can lead to the quantum phase transitions (See
Ref.\,\cite{Verrucchi}). Indeed, it has been reported that the
quantum phase transition in entanglement of two or more qubits have
been observed at finite temperature, for example, in
Refs.\,\cite{Werlang2,ChaoLi,Werlang1,Werlang3,Rong}. Therefore, in
this study, the critical point $T_{c1}$ may indicates the presence
of the quantum phase transition in entanglement for AF interaction.

\begin{figure} 
\centering{{\includegraphics[width=9cm, height=8cm]{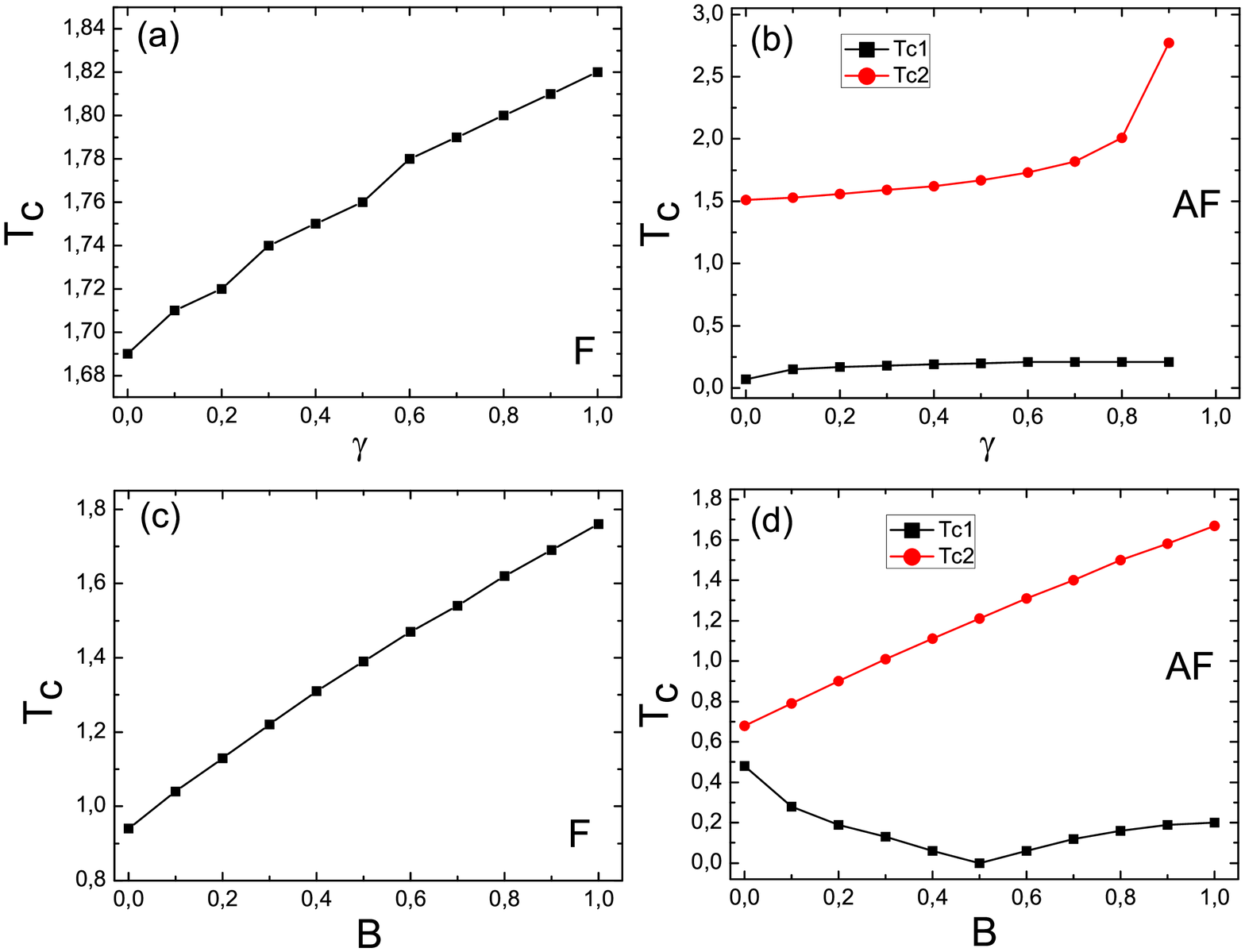}}
\caption{For fixed $T=0.01$, $B=0.5$ and $b=0.5$, the $N_{12}$ as a
function of the anisotropy $\gamma$ in (a) F case, (b) AF case. For
fixed $T=0.01$, $\gamma=0.5$ and $b=0.5$, the $N_{12}$ as a function
of the magnetic field $B$ in (c) F case, (d) AF case. }}
\end{figure}

Finally, in order to see the anisotropy $\gamma$ and magnetic field
$B$ dependence of the critical temperature, the $T_{c}$-$\gamma$ and
$T_{c}$-$B$ phase diagrams for F and AF cases in Fig.\,5 (a)-(b) and
in Fig.\,5 (c)-(d), respectively. As it can be seen in Fig.\,5 (a)
the critical point $T_{c}$ for the F case linearly changes with
anisotropy. However, unlike F case, in Fig.\,5 (b) there are two
critical point for the AF case for all anisotropy. Critical
temperature $T_{c1}$ slightly increases for small anisotropy values
but it remains almost constant for all anisotropy values. Whereas
critical temperature $T_{c2}$ smoothly increases with increasing
anisotropy, however, for large anisotropy values it increases
exponentially with increasing anisotropy. On the other hand, it can
be seen in Fig.\,5 (c) the critical point $T_{c}$ for the F case
linearly changes with external magnetic field. For AF case, in
Fig.\,5 (d) the critical temperature $T_{c1}$ linearly increases
with increasing $B$. However, the critical temperature $T_{c2}$
linearly decreases up to an critical $B$ values, and then this
critical point linearly increases with magnetic field.

In summary, in this subsection the behavior of thermal entanglement
in terms of negativity between spins (1/2,1) in three-spin
(1/2,1,1/2) anisotropic Heisenberg XXZ system depends on Hamiltonian
parameters are numerically investigated and results are shown.

\subsection{Thermal Entanglement between spins-(1/2,1/2) in spin-(1/2,1,1/2) system}

To measure entanglement between spins (1/2,1/2) in three-spin
(1/2,1,1/2) anisotropic Heisenberg XXZ system under an inhomogeneous
magnetic field for F and AF interactions, the negativity $N_{13}$ in
Eq.\,(7) is obtained after computing numerically elements of the
matrix $\rho _{13}^{t_{1}}$. Obtained results are present below.

The magnetic field dependence of the the negativity $N_{13}$ for
stagger magnetic fields $b=0$ and $b=0.5$ at fixed anisotropy
parameter $\gamma=0.5$ and temperature $T=0.01$ are shown for F case
$J=-1$ and AF case $J=1$ in Fig.\,6 (a) and (b), respectively As it
can be seen from Fig.\,6 that typically two plateau appear in
negativity curve in the case of F, however, only one plateau appears
for AF interactions with two critical points. Transitions between
plateaus are quite straight and they occur at critical $B_{c1}$ and
$B_{c2}$ as well in Fig.\,2. However there is no a singularity at
critical $B_{c1}$ in $N_{13}$ curves unlike $N_{12}$. For F case,
first plateau in Fig.\,6 (a) appears up to critical point $B_{c1}$
then second plateau occurs between $B_{c1}$ and $B_{c2}$ where
entanglement drops to zero. For AF case, nonzero plateau only
appears between  $B_{c1}$ and $B_{c2}$. Here also the stagger
magnetic field does not play role as a plateau mechanism however it
changes the width and height of the entanglement plateaus.

\begin{figure} 
\centering{{\includegraphics[width=9cm, height=6cm]{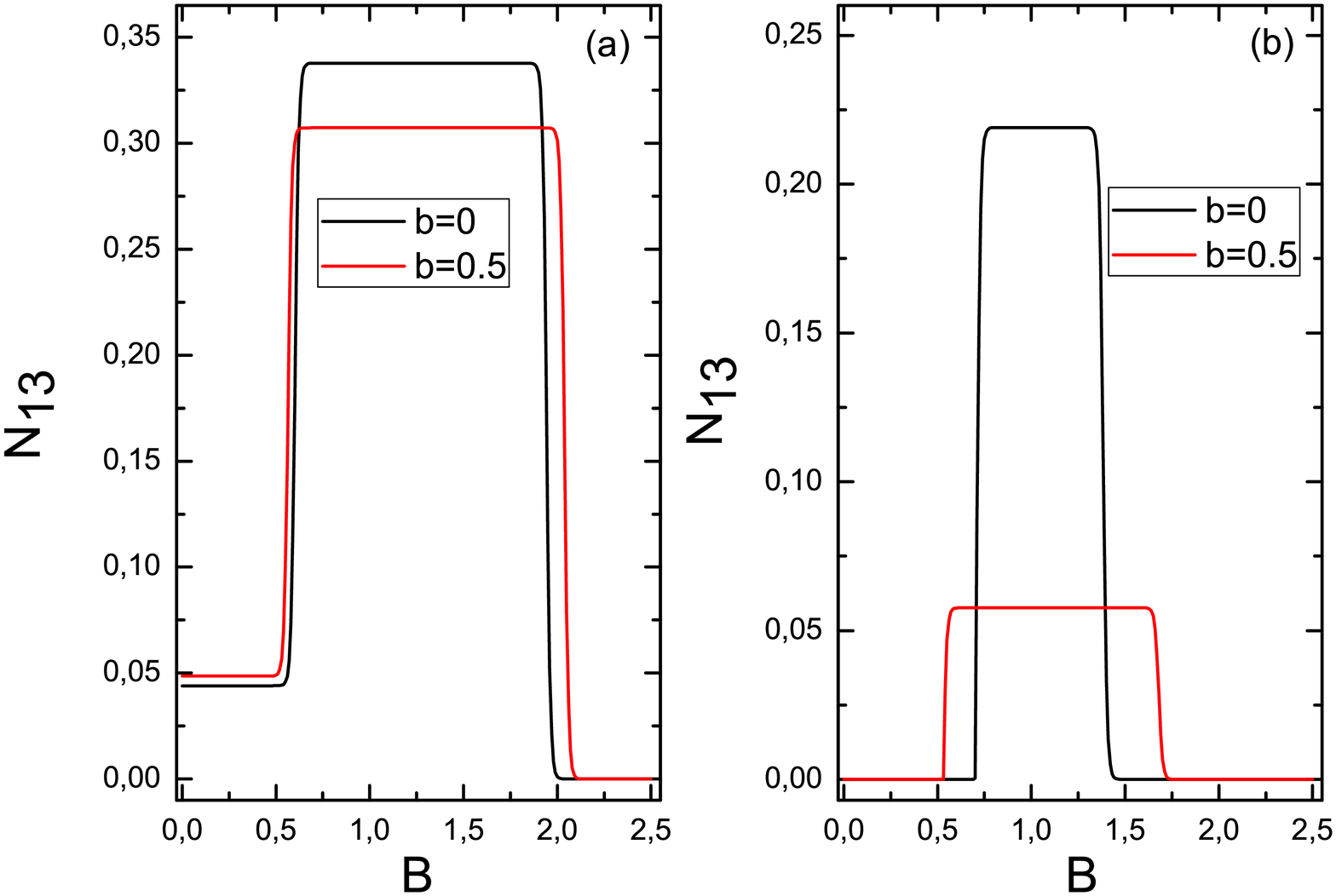}}
\caption{Magnetic field dependence of the $N_{13}$ for arbitrary
anisotropy and staggered magnetic field values at very low
temperature (a) F case $J=-1$ (b) AF case $J=1$. In (a) and (b) the
parameters are set as $\gamma=0.5$, $T=0.01$, $b=0$ and $b=0.5$. }}
\end{figure}

In order to see effect of the anisotropy on the critical point
$B_{c}$ in the case of F and AF, Fig.\,7 is plotted for fixed
temperature and staggered field values. As it can be seen from
Fig.\,7 (a) and (b) that critical $B_{c1}$ and $B_{c2}$ linearly
increase with increasing anisotropy $\gamma$ value for fixed
temperature and staggered field values. This behavior indicates that
the width of the plateaus also linearly decreases when anisotropy
decreases from one to zero. Similarly for $\gamma>0$, three-spin
(1/2,1,1/2) F and AF triangle behaves as XXZ system and when
$\gamma=0$ the system reduces to XX system. Hence on can conclude
that plateau behavior in entanglement between (1/2,1/2) spins occurs
in three-spin (1/2,1,1/2) XX Heisenberg spin system.
\begin{figure} 
\centering{{\includegraphics[width=9cm, height=6cm]{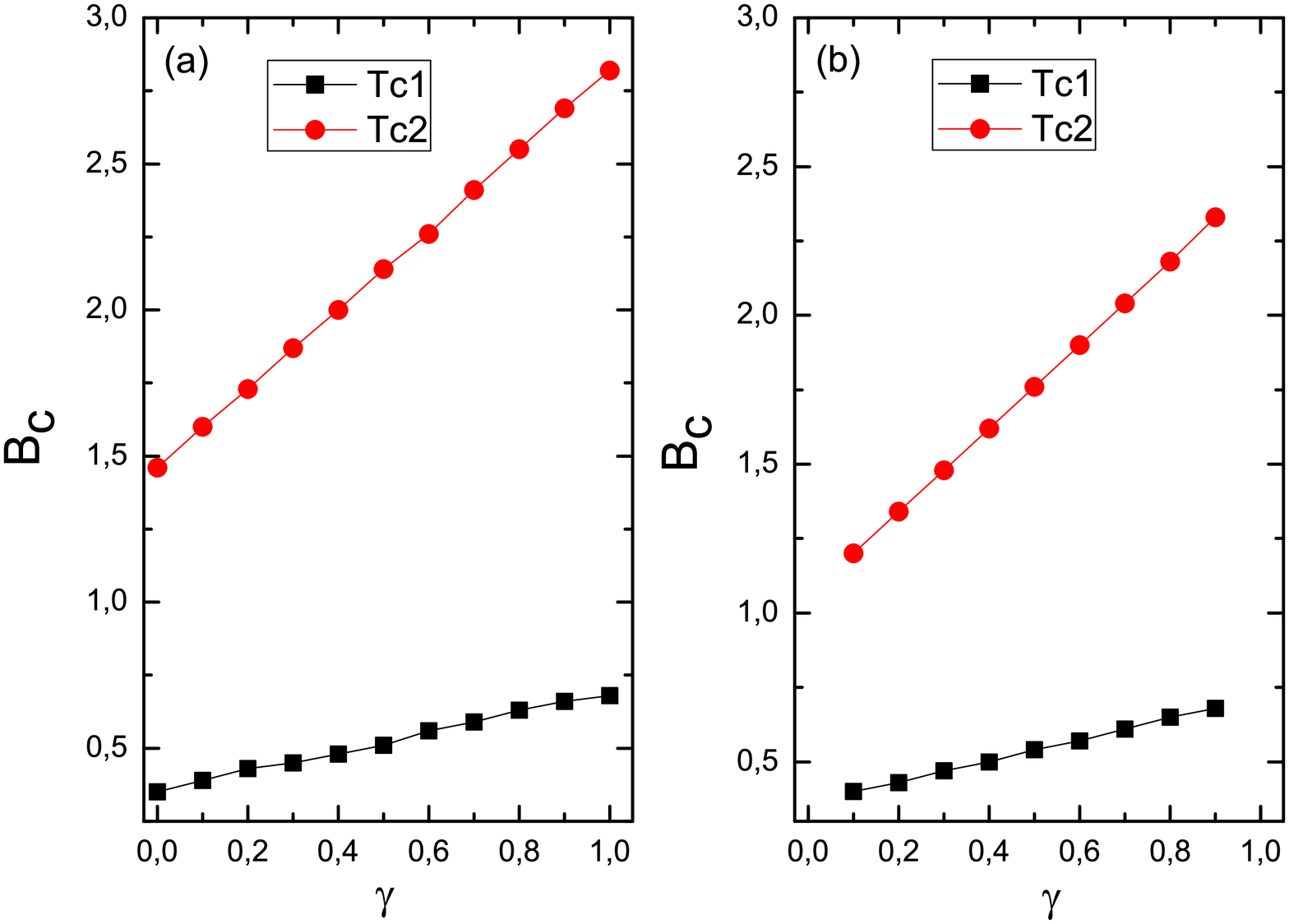}}
\caption{Anisotropy dependence of the $N_{13}$ in the F and AF XXZ
systems for $B=1$, $b=0.5$ and $T=0.01$ in (a) F case ie., $J=-1$,
(b) AF case i.e., $J=1$.}}
\end{figure}

Temperature dependence of the negativity $N_{13}$ for F and AF cases
at fixed $J=\pm1$, $\gamma=0.5$, $B=1$ and $b=0.5$ values is in
Fig.\,8. It can be seen from figure that thermal entanglement in F
case is more sensitive to the temperature than that in AF. The
negativity $N_{13}$ decreases monotonically with temperature $T$ and
finally reaches to zero at a critical temperature $T_{c}$ for both
of F and AF cases. The value critical point $T_{c}$ for AF case is
smaller than that of F case. On the other hand, it doesn't seem
trace of quantum phase transition in temperature dependence of the
$N_{13}$ unlike $N_{12}$.
\begin{figure} 
\centering{{\includegraphics[width=8cm, height=6cm]{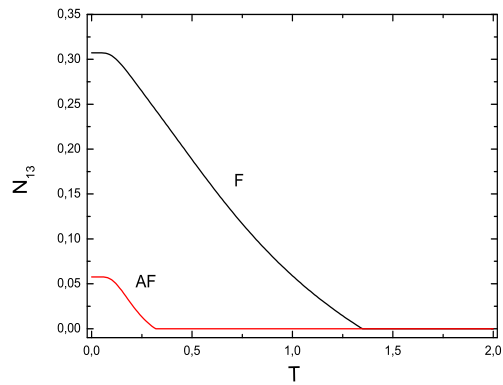}}
\caption{The $N_{13}$ as a function of temperature $T$ at $B=1$,
$b=0.5$ and $\gamma=0.5$ for F case ie., $J=-1$ and AF case i.e.,
$J=1$.}}
\end{figure}

Finally, the $T_{c}$-$\gamma$ and $T_{c}$-$B$ phase diagrams for F
and AF cases in Fig.\,9 (a)-(b) and in Fig.\,9 (c)-(d),
respectively. These phase diagrams present the anisotropy $\gamma$
and magnetic field $B$ dependence of the critical temperature. As it
can be seen in Fig.\,9 (a) firstly critical point $T_{c}$ has a
constant value while increasing anisotropy up to critical anisotropy
$\gamma_{c}$, however, after this critical point $T_{c}$ for the F
case linearly changes with anisotropy. Correspondingly, unlike F
case, in Fig.\,9 (b) the critical temperature $T_{c}$ slightly
increases up to critical anisotropy $\gamma_{c}$ and then passing
through a maximum value it almost linearly decreases with increasing
anisotropy. On the other hand, as it can be seen in Fig.\,9 (c)
critical temperature $T_{c}$ linearly increases with increases $B$
in the case of F. Then, in the case AF, critical temperature $T_{c}$
has zero value up to a critical magnetic field $B_{c}$, however,
after this critical point, $T_{c}$ linearly increases with magnetic
field $B$.
\begin{figure} 
\centering{{\includegraphics[width=9cm, height=8cm]{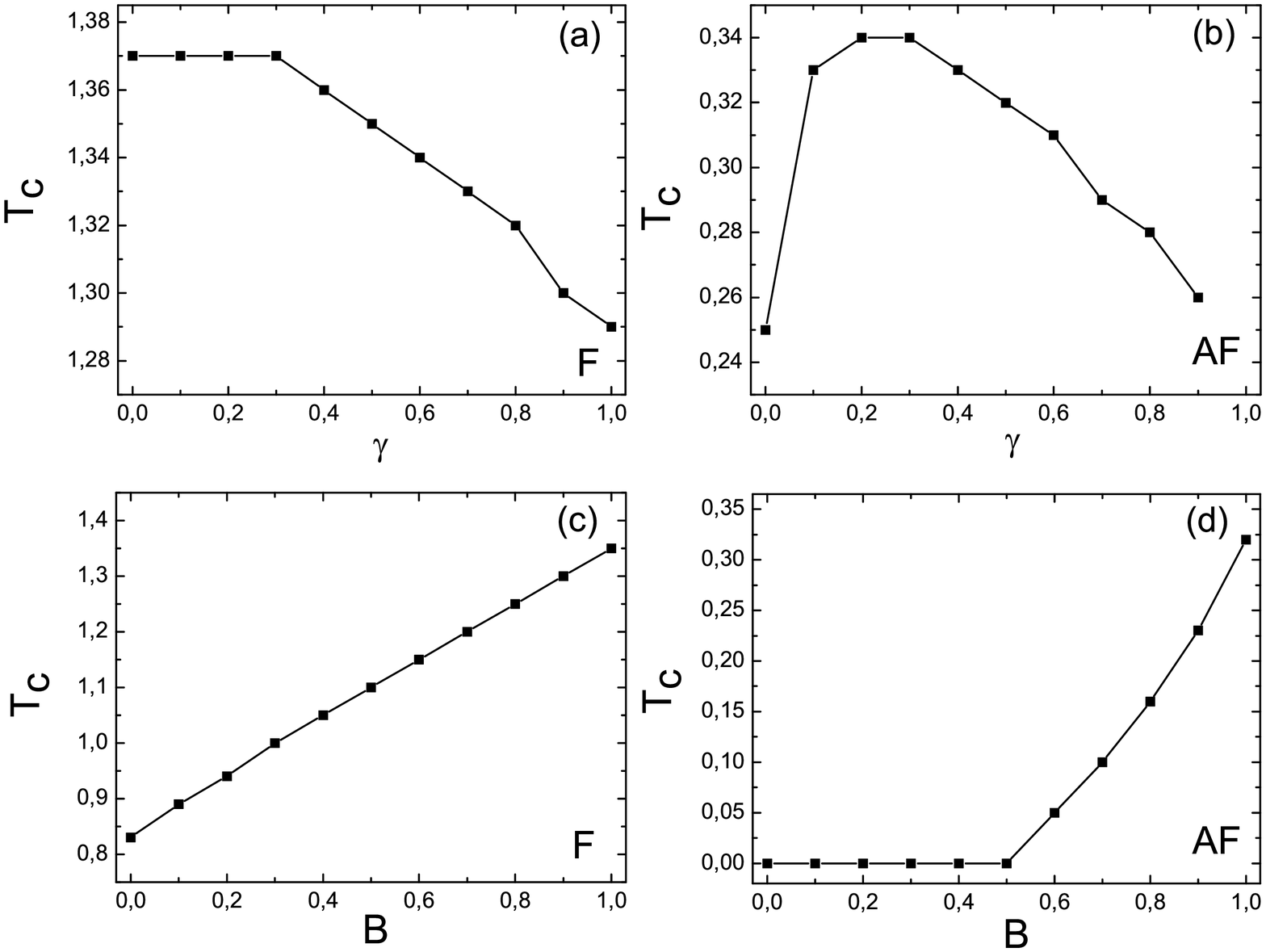}}
\caption{For fixed $T=0.01$, $B=0.5$ and $b=0.5$, the $N_{13}$ as a
function of the anisotropy $\gamma$ in (a) F case, (b) AF case. For
fixed $T=0.01$, $\gamma=0.5$ and $b=0.5$, the $N_{13}$ as a function
of the magnetic field $B$ in (c) F case, (d) AF case. }}
\end{figure}

In summary, in this subsection, the thermal entanglement between
spins (1/2,1/2) in three-spin (1/2,1,1/2) anisotropic Heisenberg XXZ
system depends on Hamiltonian parameters are numerically
investigated in terms of negativity and obtained results are
presented.

\section{Conclusion}

In this study, considering three-mixed-spin (1/2,1,1/2) anisotropic
Heisenberg XXZ system on a simple triangular cell under an
inhomogeneous magnetic field at equilibrium we numerically
investigate thermal pairwise entanglement in terms of the negativity
of the spins (1/2,1) and (1/2,1/2) depends on parameters of the
Hamiltonian (1).

We show that different and strong entanglement plateau formation in
negativity curves  $N_{12}$ and  $N_{13}$ which are respectively
correspond to thermal entanglement of the spin pairs (1/2,1) and
(1/2,1/2) occurs at depends on external magnetic field for fixed
parameters of Hamiltonian for both of F and AF interactions.
Numerical results reveals that the appearing or disappearing of the
entanglement plateaus and transition between plateaus occur at
different critical magnetic values. We see that these critical
magnetic values, on the other hand, height and width of the plateaus
are affected by anisotropy, stagger magnetic field and interaction
type of Hamiltonian (1). We conclude that plateau behaviors in
thermal entanglement curves are caused by the energy gap between
singlet ground states and the first excited triplet states. On the
other hand, we also separately investigate temperature dependence of
$N_{12}$ and $N_{13}$ for both of F and AF cases. We see that the
negativities $N_{12}$ for only F case and $N_{13}$ for both F and AF
cases almost decays exponentially to zero at an critical temperature
with increasing temperature. However, surprisingly, we observe two
critical points $T_{c1}$ and $T_{c2}$ in entanglement curve of
$N_{12}$ for only AF case. We conclude that the first critical point
$T_{c1}$ in entanglement curve of $N_{12}$ for AF case may be
signature the presence of the quantum phase transition at finite
temperature in the triangular cell with three-mixed XXZ Heisenberg
spin.

\section*{Acknowledgements}
This work is a part of MSc thesis of S. Deniz Han. Authors
acknowledge to Ozgur E. Mustecaplioglu for valuable comments and
also one of the authors (SDH) thanks to Tugba Tufekci-Sen for her
kind assistance with this work.

\end{CJK*}  
\end{document}